# Molecular Dynamics Study of Argon Flow in a Carbon Nanotube

Anuj Chaudhri

**Advisor:** Dr. Jennifer R. Lukes

# ABSTRACT

This study focuses on trying to understand the flow of argon inside carbon nanotubes. The methodology of molecular dynamics and its implementation as a tool to effectively model fluid flows inside nanocomponents is established, followed by an understanding of the intermolecular potentials which effectively model the interactions between the argon and carbon atoms. Argon is one of the most non-reactive elements in the periodic table. Being a noble gas the intermolecular interactions can be easily modeled using simple potential energy functions. The reason for using Argon lies in the simplicity in modeling its behavior and benchmarking the results with existing studies in literature. The first two cases serve to benchmark the numerical scheme by plotting the flow velocity and the structures of argon atoms while flowing inside and outside the tube. The third case simulates argon flowing from a reservoir which is twice as big as the largest dimension of the argon crystal. The simulation is carried out on short tubes with low densities of atoms. Two stages of transport are observed during the simulations. The first stage is characterized by the entry of the argon atoms into the tube which depends strongly on the repulsive forces between the argon atoms and the attractive forces between the argon and carbon tip atoms. The second stage is more chaotic as the molecules travel back and forth in the tube due to competition between the kinetic energy and the attractive potential energy of the carbon atoms on the argon. It is also observed that the argon atoms move closer to the walls of the tube. In case of higher densities they move in a coordinated fashion forming spiral-like structures. The transport is erratic and is observed in all the cases, indicating that the motion of atoms inside nanocomponents is highly randomized, yet governed solely by the intermolecular forces.

# CONTENTS

# Chapter 1
# INTRODUCTION

## 1.1 Background and scope

'Nanotechnology' or the science of manipulating atoms and molecules to make structures and devices having novel properties has received considerable attention for a few years now. Before such devices can be used it is critical to understand the fundamental laws that govern their functioning. The laws of continuum fluid mechanics are very well developed and understood. The Navier-Stokes equations are the fundamental constructs which need to be solved for a macroscopic continuum fluid flow. It would be appropriate to develop such constructs for a fluid flow at the nanoscale also. Fluid flow through nano machines or nano components would have a wide variety of applications in biotechnology and biochemical synthesis, with a wide array of potential applications in drug discovery and delivery, deposition of optically active probes in targeted locations, DNA sequencing [1], and single molecule detection [2]. There has been a lot of work on micro porous inorganic membranes [3] and zeolites for adsorption, and separation of multi-component mixtures in nano porous materials.

Simulations of thin fluid films between confined walls [4-5] indicate that crystalline or glassy order is induced across the film. The solid walls induced both in-plane order and normal order which leads to 'non-continuum' behavior of enhanced viscosity, slow relaxation times and stick-slip motion. The discovery of carbon nanotubes [6-7] has led many groups to use the hollow cylindrical cavity present in the nanotubes to check for flow of different fluids. Studies indicate that only relatively low surface tension materials can be drawn inside nanotubes [8]. This implies that typically pure metals would not be drawn into the inner cavity of nanotubes through capillarity, but water and organic solvents would.

Several studies, both experimental and computational have been carried out in this regard. *In situ* experiments of encapsulated aqueous fluids in nanotubes [9] indicate good wettability of the inner tube walls. Since experiments are in general difficult to perform at the atomic scales, several groups have studied fluid flows computationally using molecular dynamics. Hummer et. al. [10] reported pulse-like transmissions of a one-dimensionally ordered chain of water molecules through the nanotube. This was attributed to the strong hydrogen-bonding network inside the tube. Molecular diffusive flows have also been reported for methane, ethane and ethylene [11-13]. Diffusive transport is classified into normal, single-file and intermediary modes and depends strongly on the intermolecular and molecule-nanotube interactions. The diameter is also very important in determining the mode of diffusive transport. In normal mode transport, individual molecules can pass each other whereas in single-file mode the molecules cannot pass each other because of their large size in comparison to the pore diameter.

Flow of argon and helium [14] has also been simulated under various conditions. The self-diffusivities of argon and neon are one to three orders of magnitude faster in carbon nanotubes than in silicate [15]. Similar results have been reported for transport diffusivities. The fluxes of these gases are predicted to be greater through nanotube membranes than through silicate membranes of the same thickness. Simulations on argon clusters adsorbed inside and outside nanotubes [16-17] indicate localized movements of argon atoms and no liquid-like phases in the range of temperatures in which argon is a liquid.

This report focuses on the issue of simulating fluids and understanding the trends for wall-fluid behavior at the nanoscale. Since such problems are beyond the scope of continuum approaches they are studied using molecular dynamics simulations. The flow of argon is simulated to understand the transport process and methods to observe and characterize it. The geometry and structure of carbon nanotubes is first introduced, followed by the methodology of molecular dynamics. The simulation parameters, intermolecular potentials, time-stepping algorithm are briefly described. The results of the simulations are then compared to existing ones in literature with some new insights into movement of argon inside carbon nanotubes.

# Chapter 2
# CARBON NANOTUBES

## 2.1 Introduction

Carbon nanotubes are cylindrical, all-carbon nanostructures, one atom in wall thickness and tens of atoms around the circumference, with typical diameters ~ 1.4 nm. The nanotube can be thought of as a graphene sheet rolled up as a cylinder. Carbon nanotubes are usually prepared in the laboratory using an arc-discharge evaporation method [6-7]. Although multi-wall nanotubes were discovered initially, single-wall tubes were discovered later and have been the basis of many studies because of their fundamental structure. Carbon nanotubes have exhibited extraordinary electronic and thermal properties. It is reported that the thermal and electrical conductivity of carbon nanotubes are many times higher than conventional materials. The properties are attributed mainly to the helical symmetry of the carbon atoms arranged around the cylindrical structure.

## 2.2 Structure of carbon Nanotubes

The structure of carbon nanotubes can be explained in terms of its 1D unit cell, defined by the vectors $\mathbf{C_h}$ and $\mathbf{T}$ [18]. These vectors can be observed in Fig. 2.1. Carbon atoms are at the corners of each hexagon. The circumference of any carbon nanotube is expressed in terms of the chiral vector.

$$\mathbf{C}_h = na_1 + ma_2 \qquad (2.1)$$

The vectors $\mathbf{a_1}$ and $\mathbf{a_2}$ are the fundamental basis vectors of the honeycomb lattice and n, m are integers. The chiral vector connects two crystallographically equivalent sites on a 2D graphene sheet. $\mathbf{T}$ is the lattice vector of the 1D nanotube unit cell and the rotation angle and translation vectors are $\psi$ and $\tau$ respectively. The translation vector $\mathbf{T}$, is the intersection of the vector OB with the first lattice point. The unit cell of the 1D lattice is the rectangle defined by the vectors $\mathbf{C_h}$ and $\mathbf{T}$. Fig. 2.2 shows the chiral angle $\theta$ between the vector $\mathbf{C_h}$ and the zigzag direction defined by $\theta = 0$. Three distinct types of nanotube structure can be generated by rolling up the graphene sheet into a cylinder.

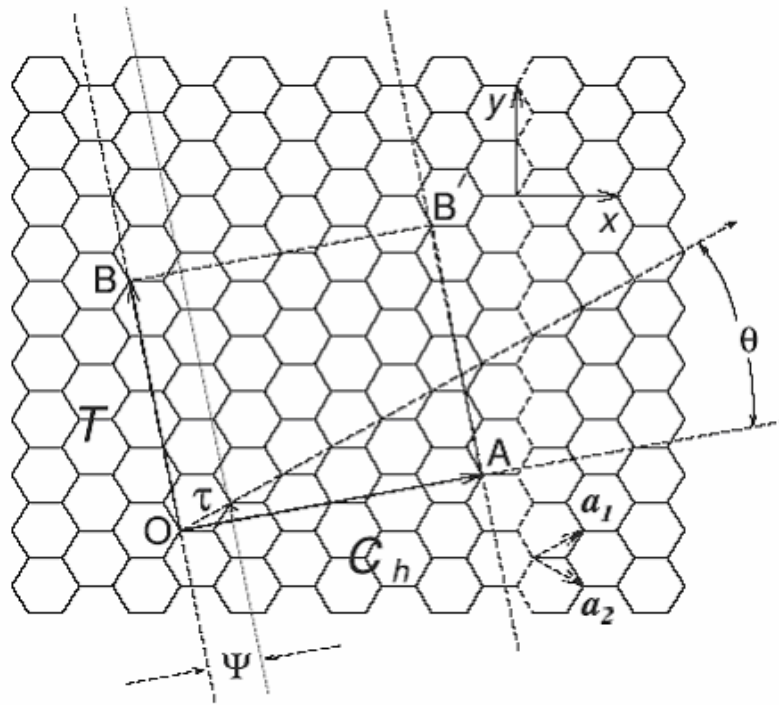

**Fig. 2.1** Schematic showing the basic unit cell structure of a carbon nanotube [18]

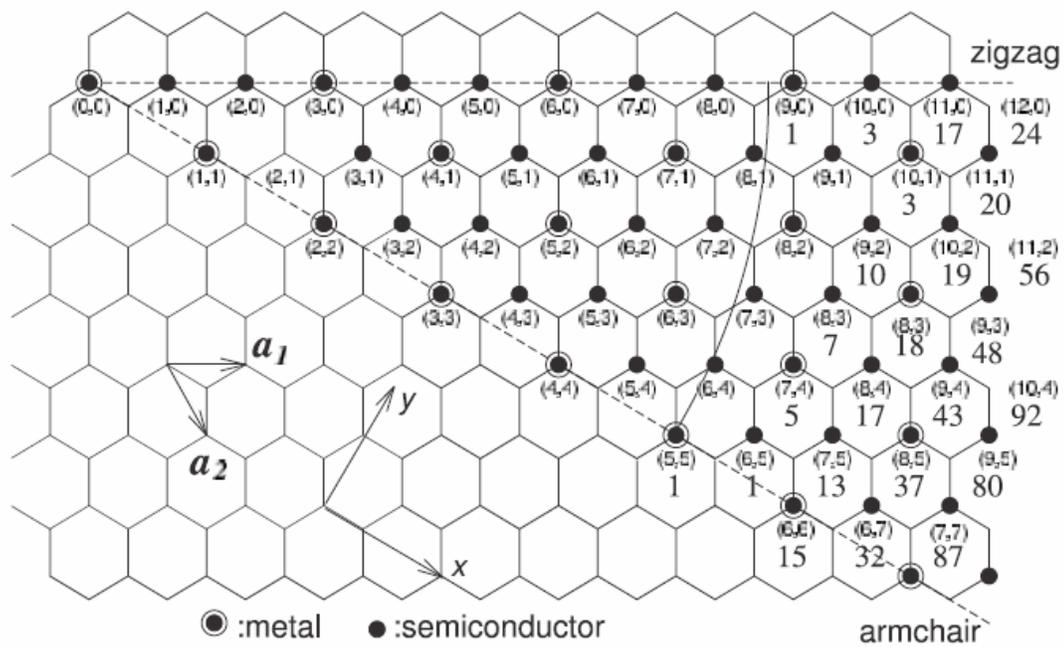

**Fig. 2.2** Vectors specified by the integers (n, m) including zigzag, armchair and chiral nanotubes. Encircled dots indicate metallic nanotubes and the others are semi conducting. [18]

The zigzag and armchair nanotubes, respectively, correspond to chiral angles of $\theta = 0$ and 30°, and chiral nanotubes correspond to $0 < \theta < 30°$. In the (n, m) notation for the type of nanotube the (n, 0) and (0, m) denote zigzag nanotubes, (n, n) denote armchair nanotubes and the (n, m) correspond to chiral nanotubes. A look at the ends of the nanotube in Fig. 2.3 shows these types of carbon nanotubes clearly.

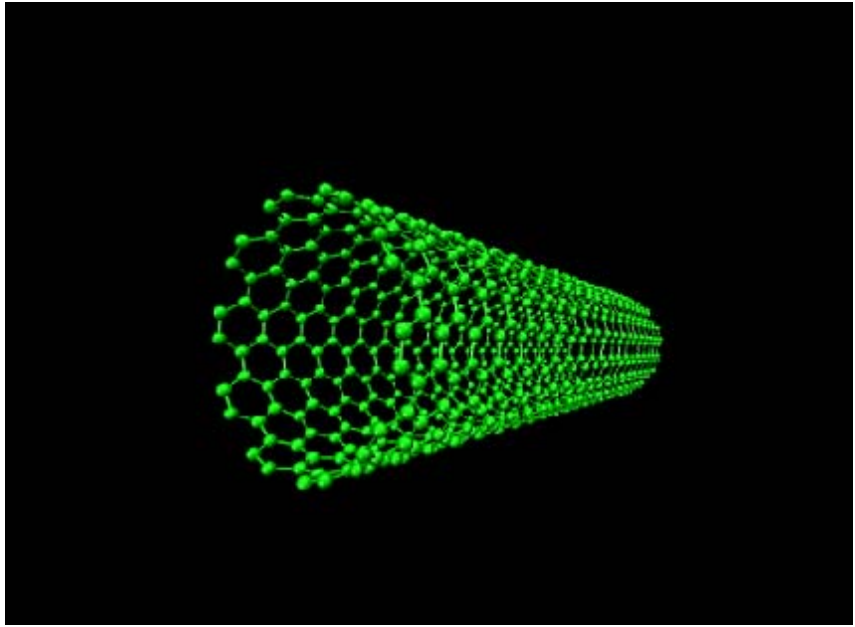

a) (10,10) armchair nanotube

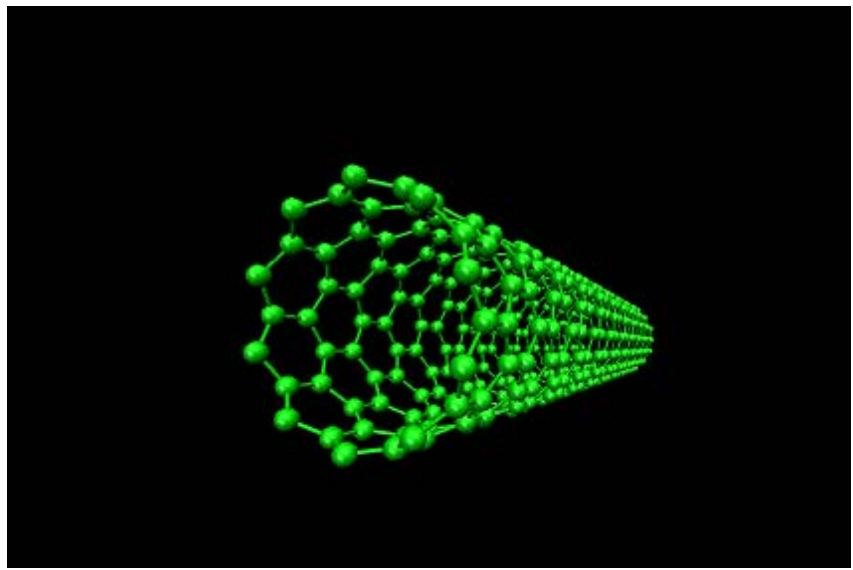

b) (10,0) zigzag nanotube

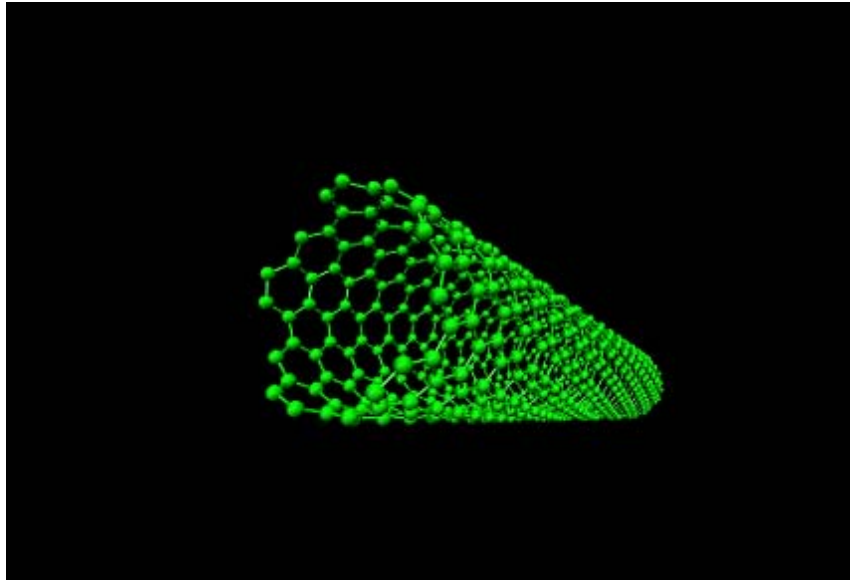

c) (10,5) chiral nanotube

**Fig. 2.3** Types of carbon nanotubes denoted by the chiral vector (n, m)

The diameter of the nanotube is given by the Eq. 2.2. The carbon-carbon bond length is taken to be on an average 1.42 A°, but it is also suggested that 1.44 A° may be a better approximation. In this report the value of 1.42 A° is used for all calculations.

$$\mathbf{d}_t = \frac{\sqrt{3}a_{C-C}(m^2 + mn + n^2)^{1/2}}{\pi} \tag{2.2}$$

Differences in the nanotube diameter $\mathbf{d}_t$ and the chiral vector $\mathbf{\theta}$ give rise to differences in the properties of the various carbon nanotubes. It is also seen that the electrical properties of the nanotube are strongly dependent on the chirality.

# Chapter 3
# MOLECULAR DYNAMICS

## 3.1 Introduction

At the onset of any simulation is a well-defined model of a physical system. The interest usually is in computing the properties of the system. The properties of the observables appear as averages over some sample space. Every system can be represented by its Hamiltonian, *H(x)*, which is a function of the generalized positions and momenta of its particles. The collection of these generalized variables constitutes a state. The set of states constitute the entire phase space $\Omega$. Hence the property of the system is also a function of the states of the system. According to Statistical Mechanics, a property *P* can be calculated if its distribution, *f (H(x))* in phase space is known.

$$<P> = \frac{\int_\Omega P(\mathbf{x}) f(H(x)) dx}{\int_\Omega f(H(x)) dx} \tag{3.1}$$

Eq. 3.1 gives the ensemble average with the quantity in the denominator called as the partition function. The distribution function specifies the appropriate ensemble for the system to be modeled. But since the ensemble is difficult to calculate in simulations the particles in the system are made to travel along a trajectory in phase space and the property calculated as a time average.

$$\bar{P}_t = (t - t_0)^{-1} \int_{t_0}^{t} P(\mathbf{x}(\tau) d\tau \tag{3.2}$$

The *ergodic* hypothesis assumes that at infinite times the above two averages would be the same. However a system cannot be made to develop over a very long time and hence the computation is carried out where the observation time is much larger than the molecular time on the assumption that,

$$\bar{P}_t \approx <P> \tag{3.3}$$

In order to propagate the system through phase space there are two main methods:

a) stochastic
b) deterministic

## 3.2 Stochastic and Deterministic methods

In stochastic methods the positions are not temporally related, but are generated stochastically by a *Markov process*, in which the current molecular configuration is only dependent on the outcome of the immediately previous configuration. The *Monte Carlo* method is a stochastic process. Such methods however suffer from the limitation that dynamical properties cannot be calculated.

In deterministic methods the time evolution of a molecular system is followed by generating the trajectories of interacting particles, by integrating the classical equations of motion. *Molecular dynamics* is one such deterministic method [19-22]. In molecular dynamics the dynamics of molecules are usually assumed to obey the classical equations of motion, which is valid provided de Broglie wavelength of the atom or molecule is much less than the average distance between the neighboring molecules, i.e.

$$\frac{h}{(3mK_bT)^{1/2}} << (V/N)^{1/2} \qquad (3.4)$$

where **h** is the Planck's constant, **m** is the atomic mass, $K_b$ is the Boltzmann constant, **T** is the temperature, and **N/V** is the molecular density.

## 3.3 Equations of motion

In molecular dynamics (MD), Newton's equations of motion are solved in time to move the atoms or molecules in phase space [23-24]. The classical equations can be written in their most basic form as,

$$\mathbf{F}_i = m_i \frac{d^2\mathbf{r}_i}{dt^2} \qquad (3.5)$$

where $m_i$ is the mass of the $i^{th}$ particle, $\mathbf{F}_i$ is the force acting on it and $\mathbf{r}_i$ in the position vector.

Molecular dynamics simulations are carried out most of the times in the NVE statistical mechanics ensemble where N-the number of atoms or molecules, V-the volume of the system and E-the energy of the system, are conserved. In such isolated systems, the system Hamiltonian is conserved, i.e.

$$\mathbf{H}(\mathbf{r}^N, \mathbf{p}^N) = \frac{1}{2m} \sum_i \mathbf{p}_i^2 + \mathbf{U}(\mathbf{r}^N) = conserved \qquad (3.6)$$

The Hamilton's equations of motion are then given by,

$$\frac{\partial \mathbf{H}}{\partial \mathbf{p}_i} = \frac{\mathbf{p}_i}{m} = \dot{\mathbf{r}}_i \qquad (3.7)$$

$$\frac{\partial \mathbf{H}}{\partial \mathbf{r}_i} = -\dot{\mathbf{p}}_i \qquad (3.8)$$

This amounts to solving 6N first order differential equations instead of the 3N second order equations in 3.5. Any conservative force can be written as the negative gradient of the potential function $\mathbf{U}(\mathbf{r}^N)$,

$$\mathbf{F}_i = \frac{\partial \mathbf{U}(\mathbf{r})}{\partial \mathbf{r}_i} \qquad (3.9)$$

The most important and time-consuming part of the simulation is the calculation of forces from the potential energy function. The potential energy function of a system of N particles can be written in terms of the coordinates $\mathbf{r}_i$ as,

$$\mathbf{U}(\mathbf{r}) = \sum_i U_1(\mathbf{r}_i) + \sum_i \sum_{j>i} U_2(\mathbf{r}_i, \mathbf{r}_j) + \sum_i \sum_{j>i} \sum_{k>j>i} U_3(\mathbf{r}_i, \mathbf{r}_j, \mathbf{r}_k) + \dots$$

(3.10)

The j>i notation indicates that summation is over distinct pairs and no pair is to be counted twice. The first term in Eq. 3.10 is the effect of an external field on the system. The second term is the pair potential which depends only on the magnitude of the pair separation $r_{ij} = |\mathbf{r}_i - \mathbf{r}_j|$. This is the most important part of the potential energy function as many system interactions can be approximated as a pair potential. The third term and beyond represent many body interactions, which are usually neglected in simulations due to computational complexity. Also for systems such as argon, which are closed-shell the pair potential has been found to be a very good approximation.

In many situations it is necessary to control the temperature of the system. Since there is no definition of temperature in dynamics, the *equipartition* theorem from statistical mechanics is invoked, which relates the kinetic energy of the atoms or molecules with the temperature as,

$$<\frac{1}{2}m\sum_{i=1}^{N}\mathbf{v}_i^2> = \frac{3}{2}NK_bT \qquad (3.11)$$

i.e. the average kinetic energy of all the particles is equal to 3/2 times $NK_bT$.

### 3.4 Procedure

MD simulations are usually done in three stages. They are as follows:

1. **Initialization**: The first stage involves setting up the initial positions and velocities of the atoms or molecules in the system. Since Newton's equations are second order initial value ordinary differential equations, both positions and velocities need to be specified. The velocities are assigned randomly from a Maxwell-Boltzmann distribution. The net linear momentum of the system needs to be conserved. All other system parameters are also set up.
2. **Equilibration**: Since the system could start from anywhere in phase space, it is necessary to equilibrate the system by relaxing it. This involves numerically integrating the equations of motion forward in time till the potential and kinetic energies become steady and fluctuate about a mean value.
3. **Production**: Once the system is equilibrated the averaging of properties can be done by continuing the simulation further. The system is in thermodynamic equilibrium and hence the averages would not be erroneous.

The simulation details for argon flow inside carbon nanotubes will be discussed in the next section.

# Chapter 4
# ARGON FLOW IN CARBON NANOTUBE
# SIMULATION DETAILS

## 4.1 Initialization Phase

### 4.1.1 Initial Positions

The system consists of a single-walled carbon nanotube with argon atoms placed in a regular cubic lattice at one end, as shown in Fig. 4.1. Different lengths and diameters of carbon nanotubes were used for the simulations. The argon atoms were started on a regular cubic lattice and were made to equilibrate to room temperatures. In some cases the argon atoms were placed inside the nanotube to understand the development of flow characteristics.

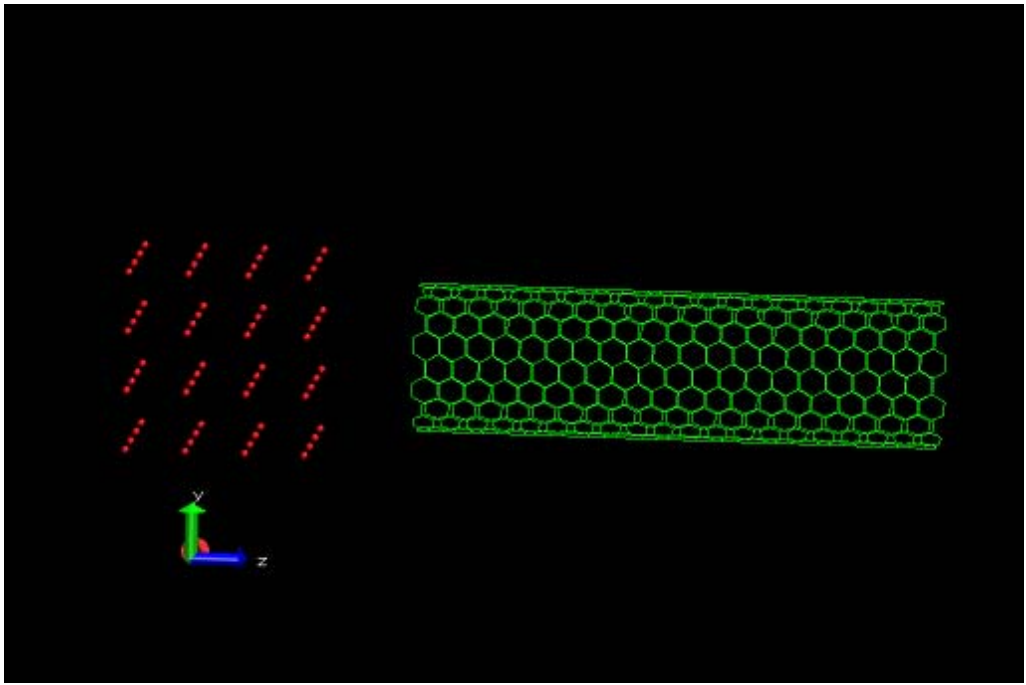

**Fig. 4.1** Initial configuration of argon atoms and carbon nanotube

### 4.1.2 Initial velocities

Random initial velocities were assigned to both argon and nanotube atoms drawn from a Maxwell-Boltzmann distribution. The velocities were scaled to the required temperatures as per Eq. 4.1.

$$T = \frac{<\frac{1}{2}m\sum_{i=1}^{N}\mathbf{v}_i^2>}{\frac{3}{2}NK_b} \tag{4.1}$$

The velocities of the carbon atoms are adjusted such that the net linear and angular momentum are zero. In other words the velocity of center of mass of the nanotube should be zero.

$$<\sum_{i=1}^{N}\mathbf{V}_i> = \mathbf{V}_{CM} = 0 \tag{4.2}$$

## 4.2 Equilibration Phase

### 4.2.1 Interatomic potential functions

The interactions between the atoms are the most important part of any MD simulation, since these would make the atoms move. The interactions between the argon atoms were simulated using the Lennard Jones 12-6 potential, which is a well-known potential used for studying argon [14-17]. The interactions between the carbon atoms are more complicated and are based on the Reactive Empirical Bond Order Potential [25-26] developed by Donald W. Brenner. The interactions between argon and carbon were modeled using Lennard Jones using different parameters. The details on these potentials will be discussed in the sections below.

#### 4.2.1.1 Lennard Jones potential

The Lennard Jones intermolecular potential is used to model interactions which are non-bonded. The origin of the Lennard Jones potential can be traced back to the van der Waal's forces which describe the interactions between fluctuating instantaneous and induced dipoles. Atoms of monoatomic noble gases such as argon and Neon interact through such type of forces.

The commonly used functional form of the Lennard Jones potential is given by Eq. 4.3. This 12-6 potential has a hard repulsive wall which arises from the Pauli's exclusion principle. When two atoms come too close to each other they atoms could undergo some 'deformation' up to a limit and then repel strongly in order to prevent the electrons from occupying the same energy levels. The Lennard Jones potential however is purely classical and it is assumed that the electrons are strongly bound to the ion cores. It serves as an approximate but highly accurate model to the real situation.

$$\mathbf{U(r)} = 4\varepsilon \left[ \left( \frac{\sigma}{\mathbf{r}} \right)^{12} - \left( \frac{\sigma}{\mathbf{r}} \right)^{6} \right] \qquad (4.3)$$

where **r** is the relative distance between two atoms. Hence the Lennard Jones potential is a pair potential describing the interactions between two atoms or molecules. $\sigma$ is the minimum distance between two atoms. As seen in Fig. 4.2 it is the point where the potential energy goes to zero. $\varepsilon$ is the energy at the minimum of the potential well.

The $-1/r^6$ term is the attractive part of the potential and is derived from dipole-dipole interactions. The $1/r^{12}$ term is the repulsive part and is used for mathematical and numerical simplicity. The parameters $\sigma$ and $\varepsilon$ determine the kind of system being modeled. In this study the argon-argon, argon-carbon and carbon-carbon non-bonded interactions were modeled using the 12-6 Lennard Jones potential.

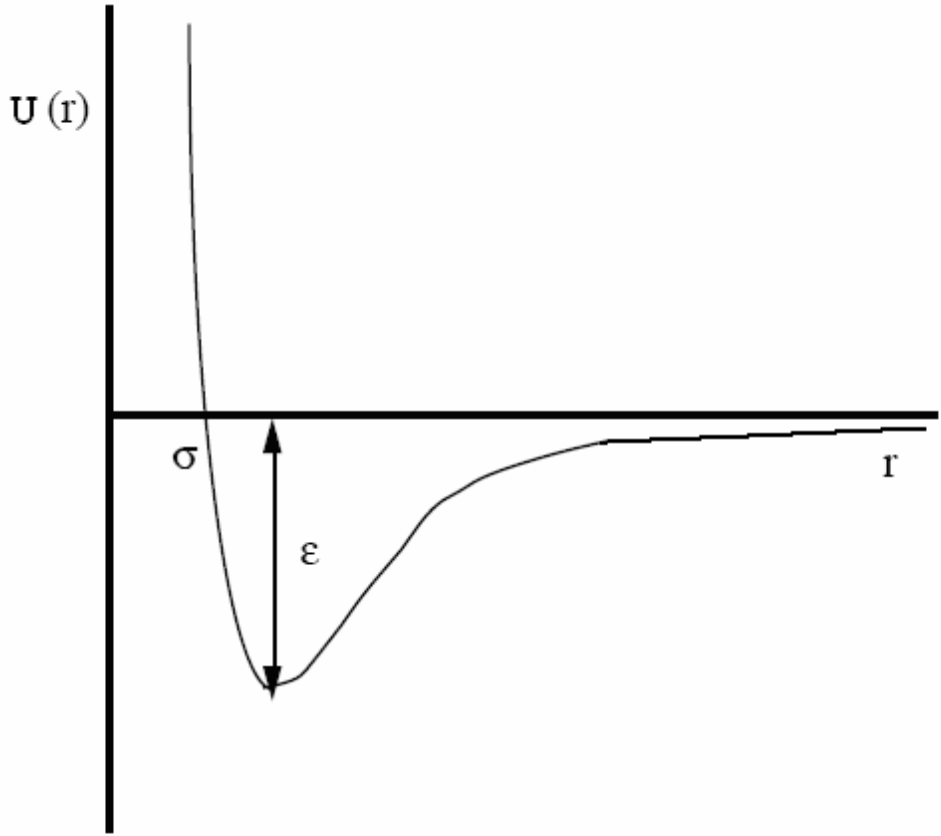

**Fig. 4.2** Shape of the Lennard Jones potential

The parameters used for the argon-argon and carbon-carbon interactions are tabulated in Table 4.1. The parameters for the argon-carbon interactions were calculated using the Lorentz-Berthelot combining rules [15, 24], i.e.

$$\sigma_{ij} = \frac{\sigma_i + \sigma_j}{2} \tag{4.4}$$

$$\varepsilon_{ij} = \sqrt{\varepsilon_i \varepsilon_j} \tag{4.5}$$

The argon-carbon interactions were truncated at 15 times the σ value, as beyond this distance the forces would be negligible.

**Table 4.1**   Lennard Jones parameters

|  | σ (A°) | ε (K) |
|---|---|---|
| **Carbon-Carbon** | 2.28 | 51.2 |
| **Argon-Argon** | 3.41 | 119.8 |
| **Argon-Carbon** | 2.845 | 78.3 |

#### 4.2.1.2   Reactive Empirical Bond Order potential (REBO)

REBO is a potential energy function for solid carbon and hydrocarbon molecules and is based on the empirical bond order formalism. Bond order in simple terms is the difference between the number of pairs of bonding electrons and anti-bonding electrons in a molecule. Hence a single bond would have a bond order of one, a double bond two and a triple bond would have a bond order of three. Fractional bond orders are also possible in structures exhibiting resonance or conjugation.

The REBO potential allows for covalent bond breaking and forming with changes in atomic hybridization within a classical potential [25-26]. It is based on the Abell-Tersoff bond order formalism in which the many-body effects are incorporated using a parameterized bond order function which introduces chemical bonding into a pair potential. In the second generation REBO potential used in this study the bonded interactions are modeled using the original REBO whereas the non-bonded inter-molecular interactions are modeled using the Lennard Jones potential with the parameters described above. The Lennard Jones potential is switched-off at short distances when two atoms are covalently bonded. The Lennard Jones potential works only if the atoms are at short distances and are not likely to form a bond or if they are not vicinal neighbors in the same molecule [27].

The analytic form for the potential energy can be written as a sum over nearest neighbors as,

$$E_b = \sum_i \sum_{j(>i)} \left[ V^R(\mathbf{r}_{ij}) - b_{ij} V^A(\mathbf{r}_{ij}) \right] \quad (4.6)$$

The functions $V^R$ and $V^A$ are pair-additive interactions that represent all interatomic repulsions and attractions between electrons and ion cores. The term $b_{ij}$ is the bond order between atoms $i$ and $j$. The bond order depends on the environment around each atom i.e. the local coordination number $N$, and the bond angles. The forms used for $V^R$ and $V^A$ are,

$$V^R(\mathbf{r}_{ij}) = f^c(\mathbf{r}_{ij})(1 + \frac{Q}{\mathbf{r}_{ij}}) A e^{-\alpha \mathbf{r}_{ij}} \quad (4.7)$$

$$V^A(\mathbf{r}_{ij}) = f^c(\mathbf{r}_{ij}) \sum_{n=1,3} B_n e^{-\beta_n \mathbf{r}_{ij}} \quad (4.8)$$

The values *A, $B_n$, α, $β_n$* are fitted empirically. The actual bond order term $b_{ij}$ is more complex and involves the contributions of the hybridized sigma bonds and the pi bonds- formed due to the overlapping of the orbitals. The empirical bond order function can be written as,

$$\overline{b}_{ij} = \frac{1}{2}\left[ b_{ij}^{\sigma-\pi} + b_{ji}^{\sigma-\pi} \right] + b_{ij}^{\pi} \quad (4.9)$$

The values for $b_{ij}^{\sigma-\pi}$ depend on the local coordination number and bond angles for $i$ and $j$ respectively. The function $b_{ij}^{\pi}$ depends on whether a bond has radical character i.e. it determines the fraction of single bond character in a double or triple bonded system and so on, and on the dihedral angle for carbon-carbon double bonds. All the above expressions define the covalent bonding energy for any collection of Hydrogen and carbon bonds. As the local bonding environment determines the effective interatomic interactions, the REBO potential can be used accurately for many systems including carbon nanotubes.

### 4.2.2 Numerical solution of Newton's equations

The numerical algorithm for solving second-order Newton's equations must be fast, require less memory, accurate and must reproduce the classical trajectory as closely as possible. The algorithm must also be time-reversible and must permit the use of a long time step. In MD an efficient and fast means of solving the equations is to use finite differences. Since the classical trajectories are assumed to be continuous, the positions and velocities can be marched forward in time using Taylor's series. The value of the time step must be significantly smaller than the time taken for an atom or molecule to travel its own length. The typical time steps in MD range from 0.1 fs (femto seconds) to 10 fs depending on the size of the system. In this study, the time step varied from 0.1-1 fs.

The *Velocity Verlet* algorithm is used in this study. It is known to be stable, accurate, time reversible and easy to implement. It involves calculating the new positions by marching forward in time using the current velocities and accelerations or forces. The velocities at half time step are then evaluated. The new forces are calculated using the new positions followed by moving the velocities forward by a half step again. The implementation can be understood using Eqs. 4.10 - 4.14.

$$\mathbf{r}(t+\Delta t) = \mathbf{r}(t) + \mathbf{v}(t)\cdot \Delta t + \frac{1}{2}\mathbf{a}(t)\cdot \Delta t^2 \qquad (4.10)$$

$$\mathbf{v}(t+\frac{\Delta t}{2}) = \mathbf{v}(t) + \frac{1}{2}\mathbf{a}(t)\cdot \Delta t \qquad (4.11)$$

$$\mathbf{a}(t+\Delta t) = -\frac{1}{m}\nabla \mathbf{V}(\mathbf{r}(t+\Delta t)) \qquad (4.12)$$

$$\mathbf{v}(t+\Delta t) = \mathbf{v}(t+\frac{\Delta t}{2}) + \frac{1}{2}\mathbf{a}(t+\Delta t)\cdot \Delta t \qquad (4.13)$$

### 4.2.3 Equilibrium structure

The argon and the carbon nanotube atoms are made to evolve separately by solving Newton's equations in time to generating particle trajectories. During equilibration there is no interaction between the argon and carbon atoms. The structures are made to relax as per the set temperatures. Velocities are scaled periodically to maintain the temperature. In the cases where argon is simulated inside a reservoir close to one end of the nanotube, it is not allowed to escape the walls of the reservoir. When the system reaches a thermodynamic equilibrium, after periodic scaling of velocities, the simulation is stopped.

## 4.3 Production phase

The argon and carbon atoms are made to interact with each other and the particle trajectories are calculated as per the forces. The system is allowed to evolve further and the properties are estimated by averaging over time.

The next chapter looks at some of the simulation results for different cases of argon flow in different nanotubes.

# Chapter 5
# ARGON FLOW IN CARBON NANOTUBE
# RESULTS

## 5.1 Introduction

In this study the argon flow inside carbon nanotubes is analyzed to understand trends in fluid and wall behavior. In the first case, argon atoms start on a regular grid inside the nanotube and are given an additional velocity in the z-direction, i.e. along the length of the nanotube. The development of the velocity in the z-direction is followed and the trends are compared to existing ones in literature. The simulations were carried out for different densities of atoms and with different initial velocities.

In the second case argon starts in a regular cubic grid outside the tube and is simulated to be enclosed inside a high-pressure container. This is done by simulating walls around the argon lattice during equilibration and not allowing any atoms to escape. The wall-fluid interactions are modeled as $\sigma/r^2$, and has a strong repulsion once the atom comes within a distance of 1 A˚ from the wall. Once the atoms have been equilibrated to gas-like conditions, they are allowed to rush in towards the nanotube.

In the third case argon atoms start on a regular grid inside a cubical box which is twice the dimension of the argon crystal lattice having only one opening towards the nanotube. The wall-fluid collisions are such that an argon atom would be reflected specularly if it hits the wall. In this case the entry of the atoms into the nanotube is investigated. No additional velocities, apart from random thermal motion, are given to the argon atoms in cases two and three. The starting and final configurations of both the argon and carbon atoms are plotted to observe movement of atoms inside and outside the tube. The trajectory for a few selected atoms is followed over time to ascertain the path they would most likely take inside the tube.

## 5.2 Case 1: Argon confined inside the nanotube

In this study the results obtained are compared to [14], where argon atoms are placed in a regular cubic lattice structure, as shown in Figs. 5.1 and 5.2. Periodic boundary conditions are applied along the length of the nanotube with the minimum image convention. In periodic systems it is assumed that the unit cell is repeated and hence the forces by image atoms also need to be computed. The forces computed using the Lennard Jones potential are short ranged compared to the length of the nanotube and hence only those image cells need to be considered that adjoin the nanotube. This is the minimum image convention where the force then becomes the sum of contributions of the image atoms adjacent to the primary cell.

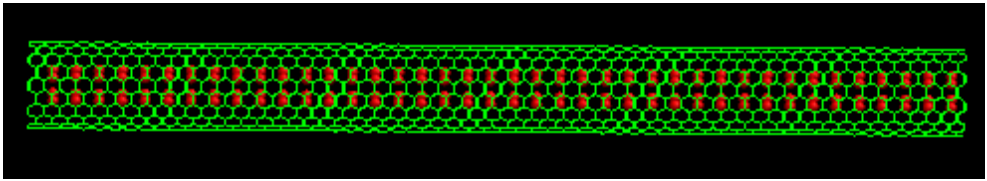

**Fig. 5.1** Transverse view of the initial argon lattice inside a 10, 10; 150 A˚ long tube

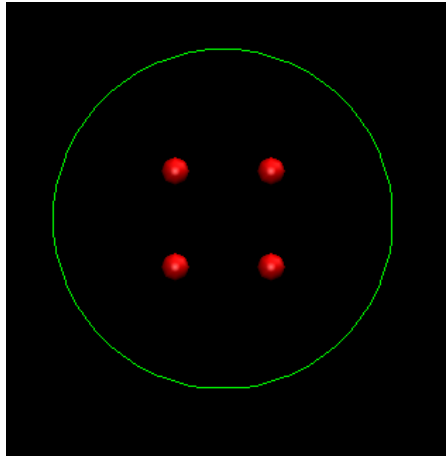

**Fig. 5.2** Axial view of the initial argon lattice inside a 10, 10; 150 A˚ long tube

The argon atoms are given initial random velocities corresponding to a temperature of 105 K. The carbon atoms are also given random thermal velocities equivalent to 350 K.

The differences in temperature arise from the amount of thermal motion that needs to be given to the atoms. Additional velocities are given to the argon atoms along the length of the nanotube. This establishes flow-like characteristics inside it. The velocities vary between 20e-04 A°/fs and 50 e-04 A°/fs. The development of the velocity component along the nanotube is plotted with time for different cases of initial velocity. Also the density of atoms inside the nanotube was varied to compare the time taken for the profiles to settle down to an average value. The number of argon atoms varied between 140 and 160. The densities correspond to cubic lattice parameters of 3.8219 A° and 3.6 A° respectively. The argon atoms were arranged such that no atom comes within 2.95 A° of any carbon atom in the tube. This is done to avoid blowing up of the atoms due to the strong Lennard Jones repulsions. The code was run for 125 ps with a time step of 0.25 fs. The system was not equilibrated as the cubical structure of the argon lattice is quickly lost within 0.125 or 0.375 ps, which is negligible compared to the total simulation time. There is no velocity scaling throughout the simulation. However the velocities of the atoms of the nanotube are conserved for linear and angular momentum, so that the fluid atoms do not move the tube in any way. All simulations correspond to the tube being dynamic i.e. the nanotube atoms are all given random thermal motion. It was ensured that the argon atoms leaving one end and entering from the other do not come too close to the other argon atoms close to the ends.

The motion of atoms is highly randomized and erratic. This has been established by the numerous films which were made to keep track of the motion of the atoms. The final configuration for the case when there are 160 argon atoms can be seen in Figs. 5.3, 5.4 and 5.5. It is observed that the atoms inside the tube shift towards the walls of the tube and arrange themselves as per the geometry of the tube. There are a few atoms in the centre at times during the simulation, but on the whole the argon atoms want to sit closer to the wall at a distance of about 2.8 A° away from the wall. This is expected, since the argon-carbon interaction is of the Lennard Jones type and 2.8 A° would correspond to the minimum distance from the wall where the potential energy would be the least. The trajectories of the atoms shown in Fig. 5.6 are plotted with respect to time in Figs. 5.7 to 5.9. It is seen clearly that the atoms tend to remain close to the tube wall. The plots also indicate the erratic motion of the atoms inside the tube, which corresponds to erratic transport and hence difficult to characterize. The films also indicate that since the argon atoms are losing energy and slowing down due to collisions, they seem to condense from gaseous to liquid phase. This is confirmed by the atoms coming closer to one another and forming clusters as seen in Fig. 5.5.

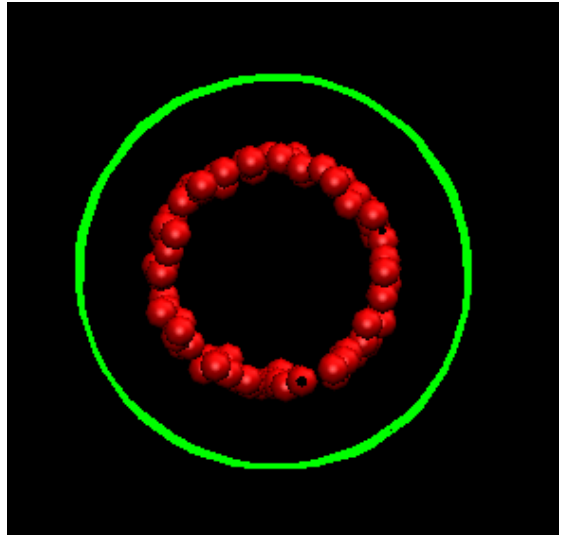

**Fig. 5.3** Final configuration of argon atoms after 125 ps

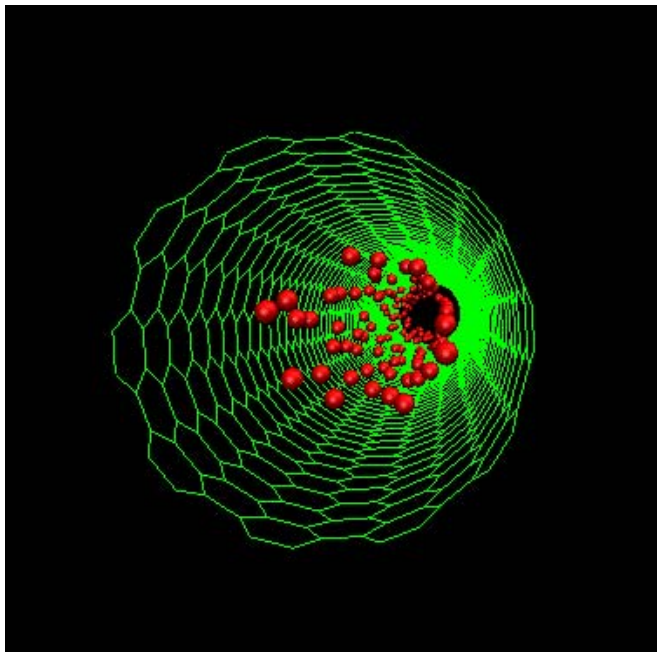

**Fig. 5.4** Another view of the final configuration of the argon atoms after 125 ps

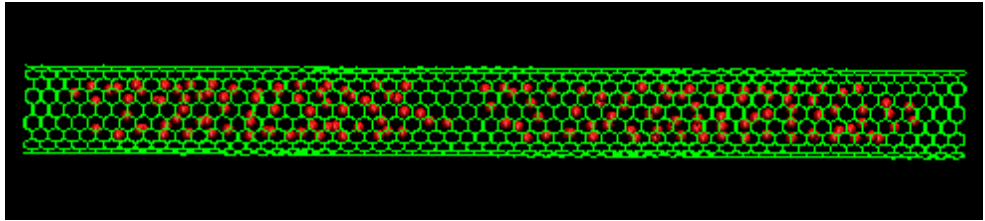

**Fig. 5.5** Snapshot showing the condensed form of argon vapor

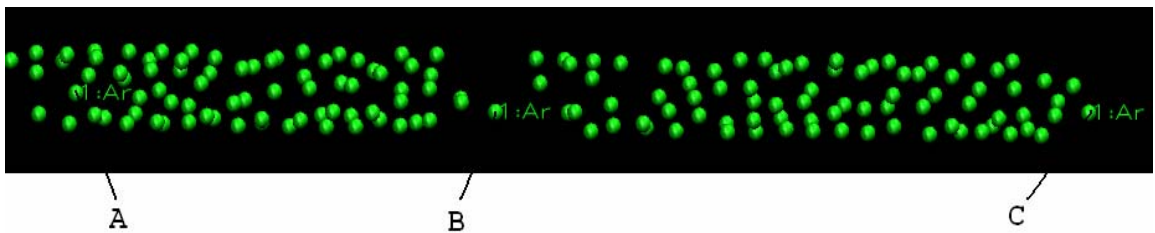

**Fig. 5.6** Randomly selected argon atoms for plotting trajectory data

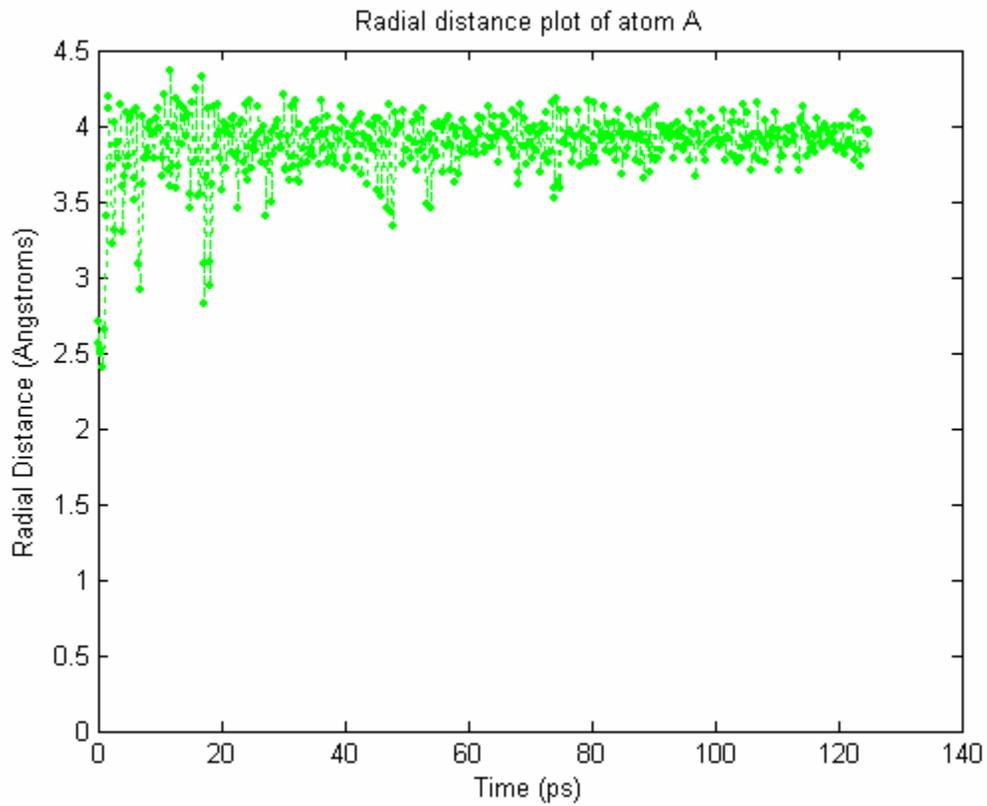

**Fig. 5.7** Radial distance plot shows that atom A stays close to the nanotube wall
Zero on the y-axis corresponds to the nanotube centre and the radius is around 6.78 A°

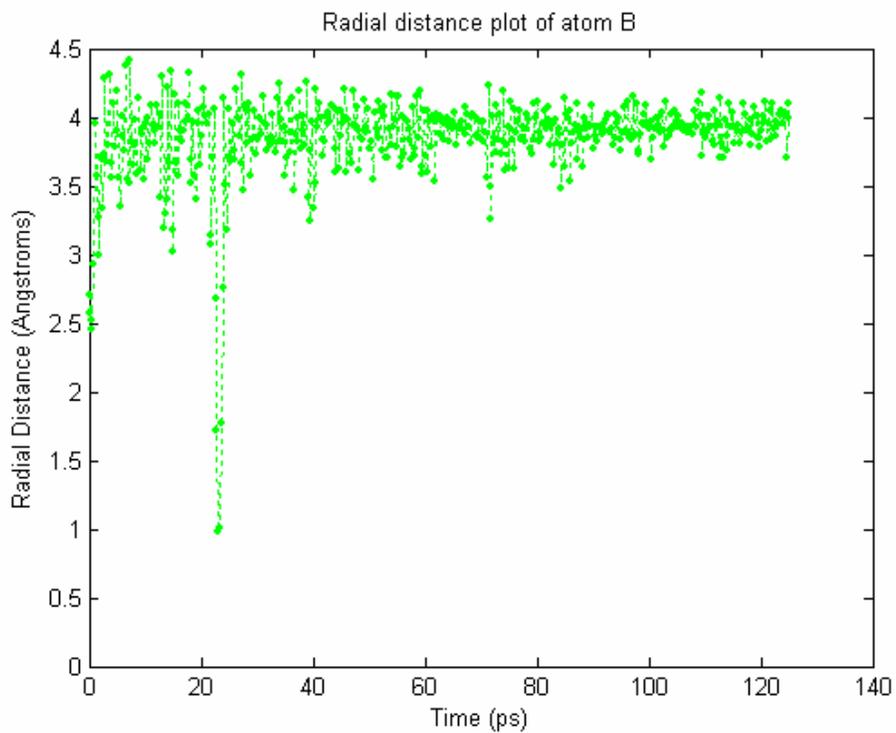

**Fig. 5.8** Radial distance plot of atom B

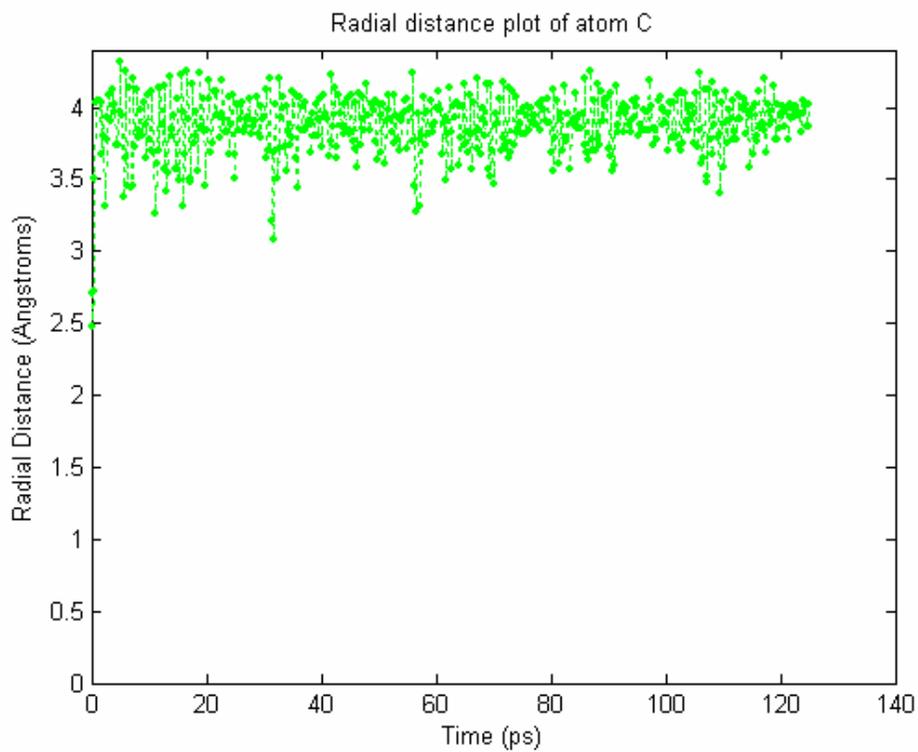

**Fig. 5.9** Radial distance plot of atom C

The above simulations were repeated for 140 argon atoms instead of 160 and for different initial velocities. The results in terms of configuration and trajectory plots look similar. The component of velocity along the length of the nanotube is plotted for the different cases in Fig. 5.10. It is seen on comparing to Fig. 5.11, that the trends look very similar.

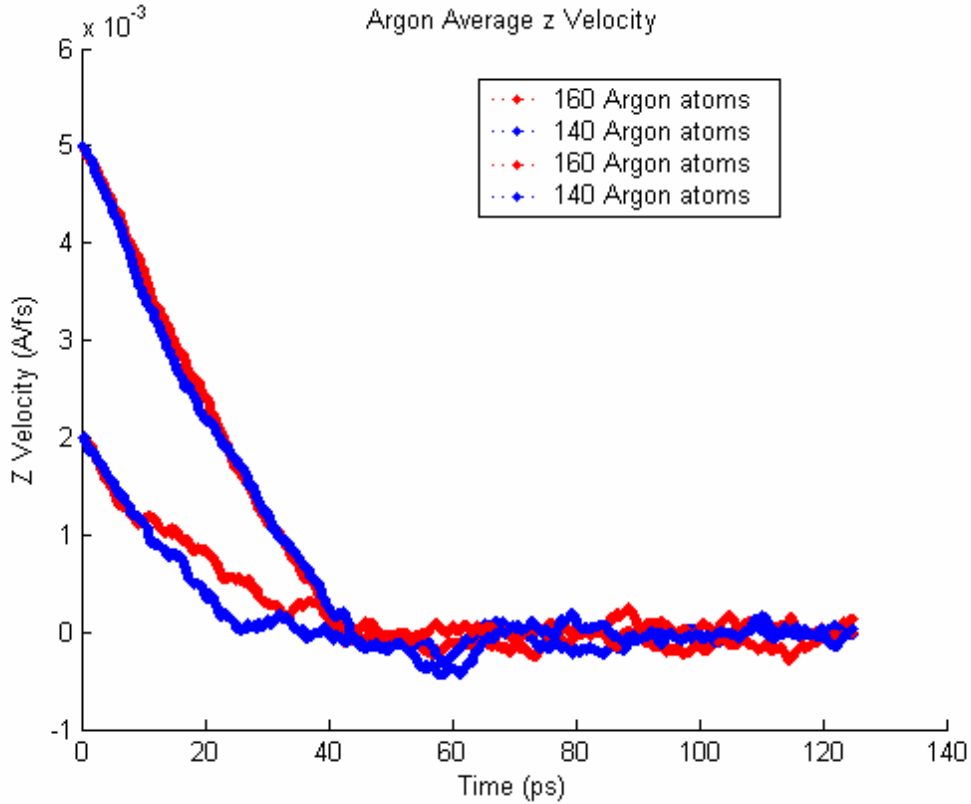

**Fig. 5.10** Average velocity profile along the length of the nanotube

It takes on average 20-40 ps for the atoms to lose the effect of the initial additional velocity and settle down. The slightly negative drift of the velocity is justifiable since once the atoms start to cool down they tend to condense and hence an apparent phase change is taking place. This causes the atoms to be pulled back to form clusters and hence the velocity component will be negative along the nanotube axis. Also for higher densities and higher velocities, the atoms seem to cool down faster as shown in Fig. 5.10. This happens since there are more fluid-fluid and fluid-wall interactions and hence the atoms lose energy faster. In the case when the velocities are lower, the higher density case takes a longer time. This could be due to the higher repulsions between argon atoms which cause the velocity to settle down slowly. From Fig. 5.11, it is seen that the order of the time taken for the velocity to settle down is the same as the above simulations. The reason for not seeing a very strong deviation in the cooling of atoms for the higher density and higher velocity case in Fig. 5.10 as compared to Fig. 5.11 is the different potentials used to characterize the interactions. The reference [14] with which some of the trends are compared also uses much higher densities and higher velocities than the cases here and this could be the reason for some of the deviations. Also the results in Fig. 5.11 are for a static tube where the atoms are not vibrating. For static tubes it takes more time for the fluid to slow down.

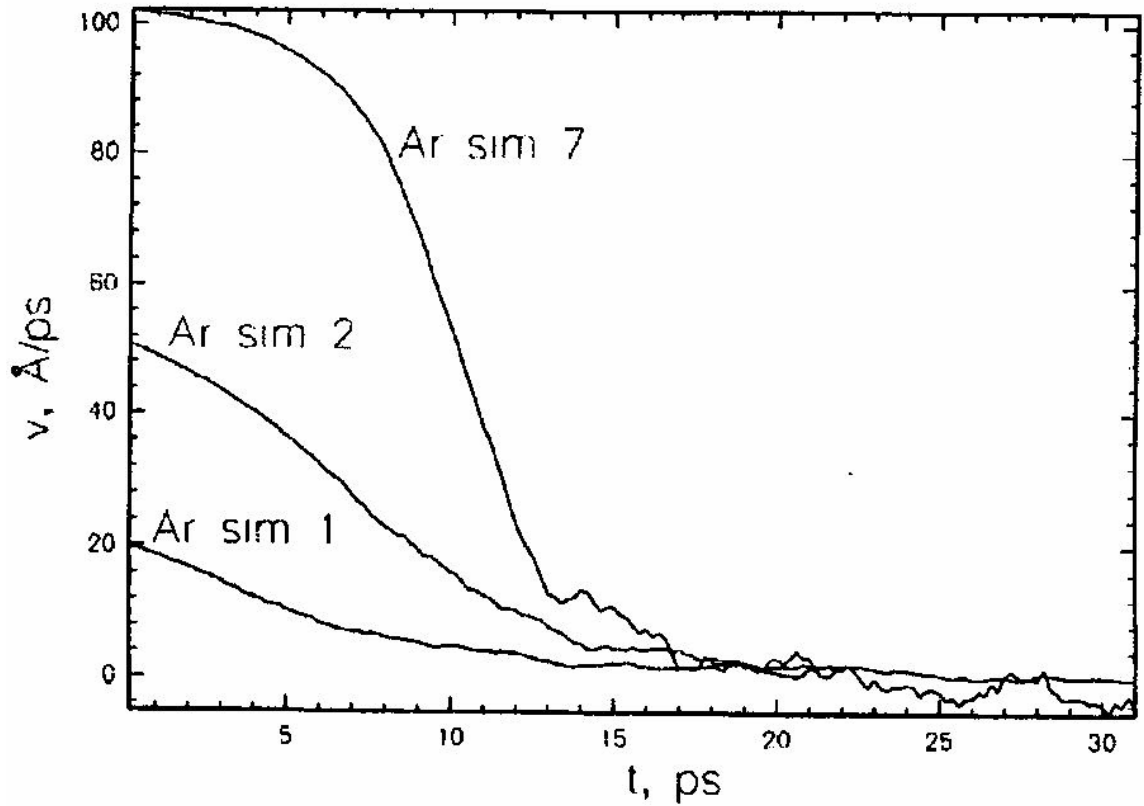

**Fig. 5.11** Average velocity profile along the length of the 10 x 10 static nanotube [14]; three different cases a.) Ar sim 1-argon temperature is 105 K, 533 argon atoms and additional z-velocity = 20.4 A°/ps, b.) Ar sim 2-argon temperature is 105 K, 533 argon atoms and additional z-velocity = 51.1 A°/ps, c.) Ar sim 7-argon temperature is 105 K, 533 argon atoms and additional z-velocity = 102.2 A°/ps

It is observed in [14] that at low fluid densities, the density profiles shift towards the tube. This is very similar to what is observed in the simulations above. Simulations were done on dynamic tubes in this study and hence it is observed that density does not have much of an effect on the velocity development. The simulations of argon flow in static tubes were not performed here since real nanocomponents would not necessarily have static tubes. Also since the diffusion length is of the order of the diameter of the tube, the development of a steady-state velocity profile is not possible and hence the self-diffusion coefficient was not calculated [14].

## 5.3 Case 2: Argon flow from a high pressure container

In this study argon starts in a regular cubic grid outside the container and is equilibrated to a steady-state temperature. The carbon nanotube is given an initial temperature of 200 K and is equilibrated separately to a steady-state temperature. The time step used for this simulation is 0.25 fs and the system is allowed to evolve for 187.5 ps. The equilibrated system is shown in Fig. 5.12. In this case the wall is modeled to have a repulsive potential and hence the argon atoms are really close to each other, giving rise to high pressures.

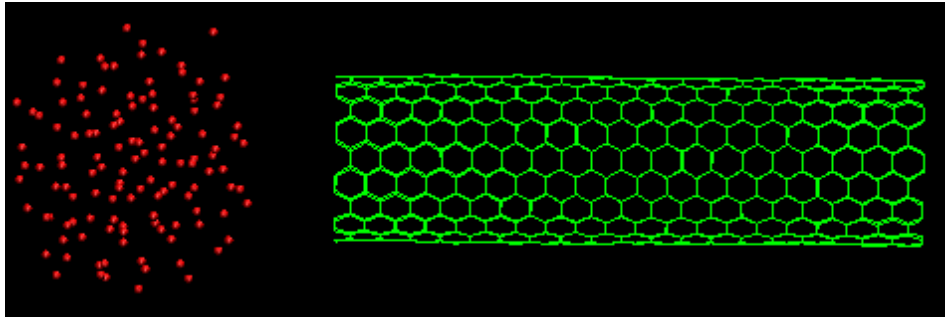

**Fig. 5.12** Equilibrated argon and carbon nanotube

After the system has been equilibrated, the entire front wall facing the nanotube is removed and the argon atoms are allowed to flow out from the container and rush towards the nanotube. It is observed that some of the atoms are lost in the process as there were no artificial walls in this case to keep them in place. The argon atoms are allowed to occupy sites both inside and outside the tube. It is observed that once argon enters the nanotube, it arranges itself in a spiral pattern close to the walls. This observation is very interesting and has been reported in some studies on argon flow inside carbon nanotubes [16-17]. The final configurations can be seen in Figs. 5.13 to 5.15. On the outside argon atoms sit in a triangular lattice that minimizes the energy of the system. This has also been independently confirmed in the studies stated earlier. It is interesting to note that even though the atoms rush towards the nanotube, most of them seem to 'stick' to the nanotube walls to form clusters. The results from references [16, 17] can be seen in Figs. 5.16 and 5.17 for comparison.

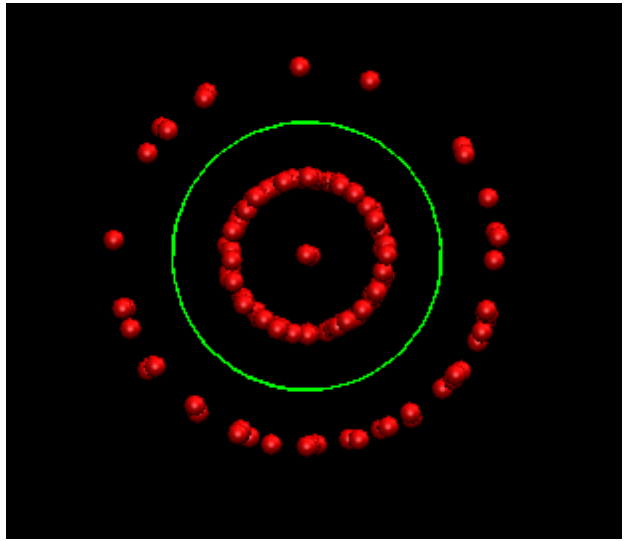

**Fig. 5.13** Axial view of the system after 187.5 ps

Though the simulations done in [16, 17] were carried out using different parameters and conditions, it is interesting to note the similarities in structures obtained once the argon atoms are in the vicinity of the nanotube. The comparison also helps to validate the numerical scheme being used for these simulations.

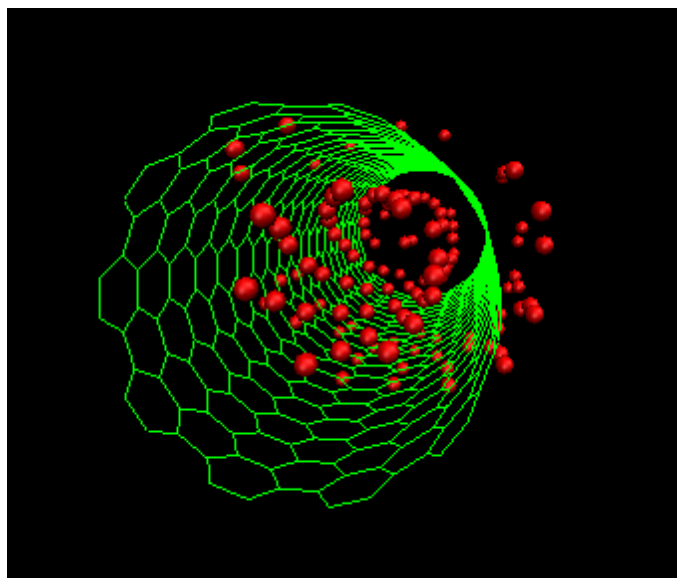

**Fig. 5.14** Argon atoms arranged in a spiral shaped pattern inside the nanotube

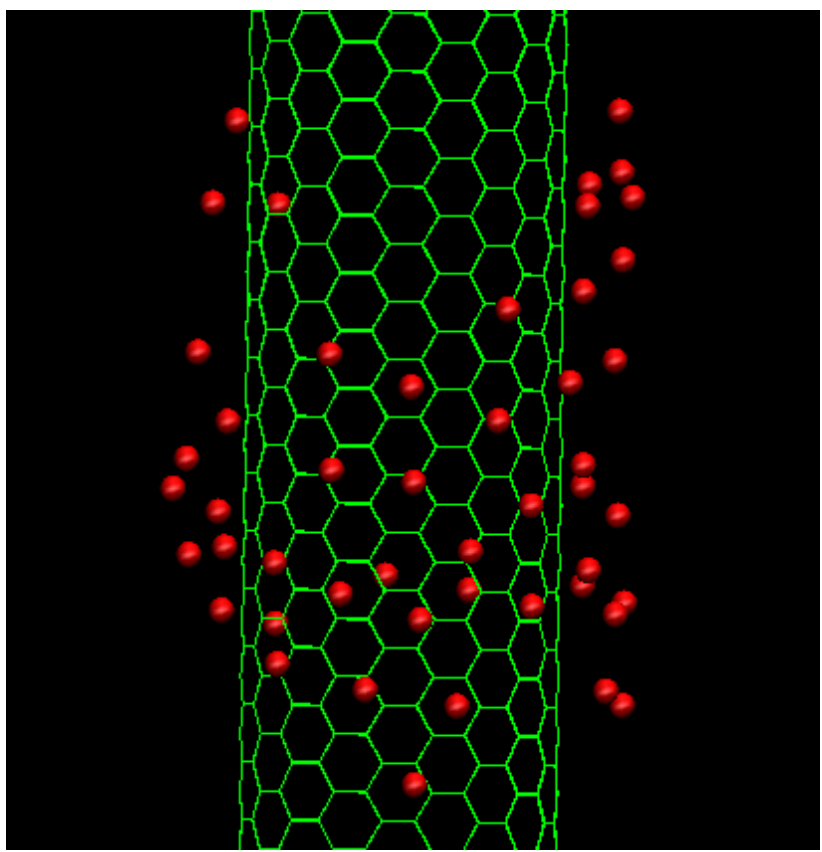

**Fig. 5.15** Argon atoms arranged in a triangular pattern outside the nanotube

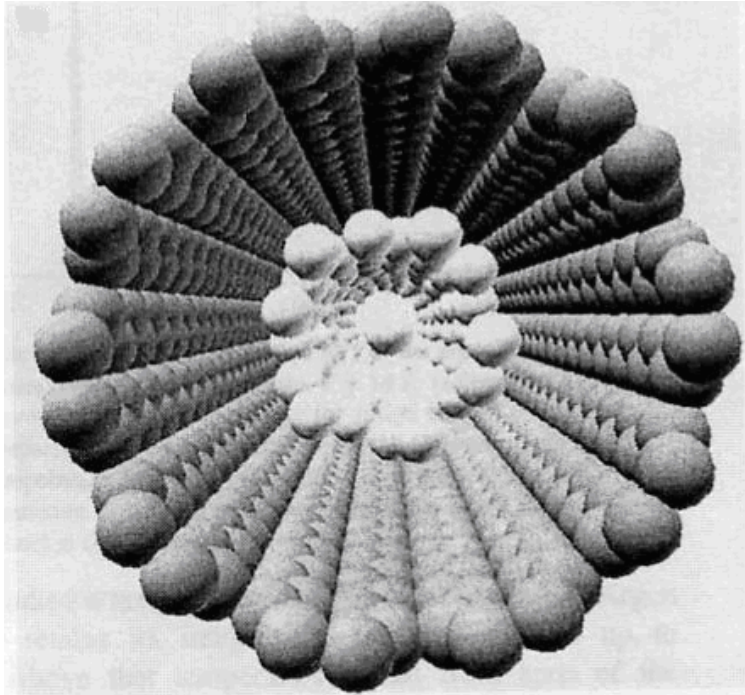

**Fig. 5.16** Argon atoms arranged inside the 10 x 10 nanotube; 110 argon atoms [16]

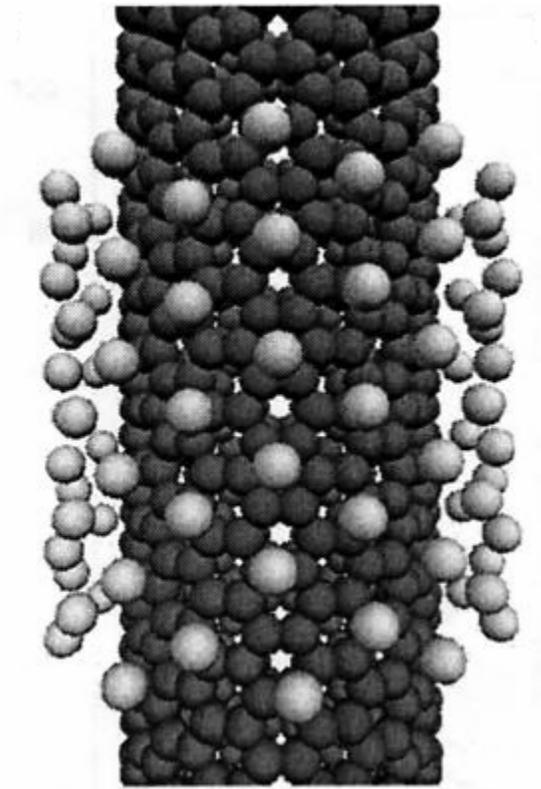

**Fig. 5.17** Argon atoms arranged outside the 10 x 10 nanotube; 110 argon atoms [17]

## 5.4 Case 3: Argon flow from a reservoir using specular wall-fluid boundary condition

In this study the argon atoms start on a cubic grid as before, but are equilibrated in a reservoir twice the size of the crystal. Both the argon and carbon atoms are given random thermal motion and the velocities are scaled periodically during equilibration to set the temperature at 300 K. During the simulation the argon atoms are only allowed to enter into the nanotube through an opening which is the size of the diameter of the tube. After traversing the length of the tube, they are allowed to leave into vacuum and not constrained any further. The number of argon atoms varied between 64 and 125. Different lengths and diameters of the tube were simulated to establish the transport during entry and exit of the fluid. Several films have been made which capture the movement of atoms from one end to another.

It was observed that the entry of the fluid atoms is completely arbitrary. The movement of the atoms inside the nanotube is completely determined by the intermolecular interactions between argon and carbon. But interestingly the argon-argon repulsion plays a vital role during the entry and exit of the fluid. To enter the tube the gaseous atom must have a larger net repulsive with another gaseous atom than with the nanotube tip atoms. This stage has been termed as non-equilibrium transport. Once the atoms have entered the tube, they will exit on the other side depending on the kinetic energy left after successive collisions with the tube walls and the potential energy between the tube atoms and the gaseous atom. The atom would be strongly affected by the attractive van der Waal's forces characterized by the Lennard Jones potential due to the argon-carbon interactions. If the gas molecule cannot overcome this force it changes direction and moves back into the tube. This is the first stage of the transport.

During the second stage the flow is more chaotic. The increase in concentration of atoms inside the tube and the movement of the atoms back and forth near the exit disturbs the movement of atoms and this initiates molecular 'bouncing' motion which has been verified by other groups working on methane and ethane diffusion in nanotubes [11-13]. As the tubes used in this study are shorter than the ones used by authors [11-13], the other stages of transport beyond the second are not observed. Also the density of the argon atoms here is very small which does not cause a buildup of density inside the nanotube and hence near steady-state transport was never seen.

The stages of transport can be understood by plotting the x, y and z trajectories of a few selected atoms over time. The simulation was carried out with 64 argon atoms using a 10, 10, 50 A˚ long tube. The time step used for this simulation is 1 fs and the simulation is run for a total time of 500 ps. Fig. 5.18 shows the plot of the x, y and z coordinates of a randomly picked atom. The plot shows clearly that the z motion of the atom (along the length of the nanotube) is very erratic with the atom moving back and forth even after it has entered the nanotube after 300 ps. The Y coordinate shown in blue indicates that the atom swings up and down the centre axis while moving forward, thus giving rise to the spiral motion. Fig. 5.19 shows the radial distance plot of the motion of the atom. As seen earlier the atom likes to stay close to the walls of the tube. Sometimes there are sudden movements close to the centre of the tube, which must be due to the competition between the kinetic energy of the atom and the repulsive potential energy near the wall. Fig. 5.20 and 5.21 show the displacement plots of another atom which crosses the entire length of the tube. It is observed that though the atom enters the nanotube later than the first on in Fig. 5.18, it exits early. Hence the second atom crosses the first one inside the tube and acquires sufficient kinetic energy to get pushed out. This indicates erratic transport of atoms inside the tube which cannot be explained or analyzed by continuum methods.

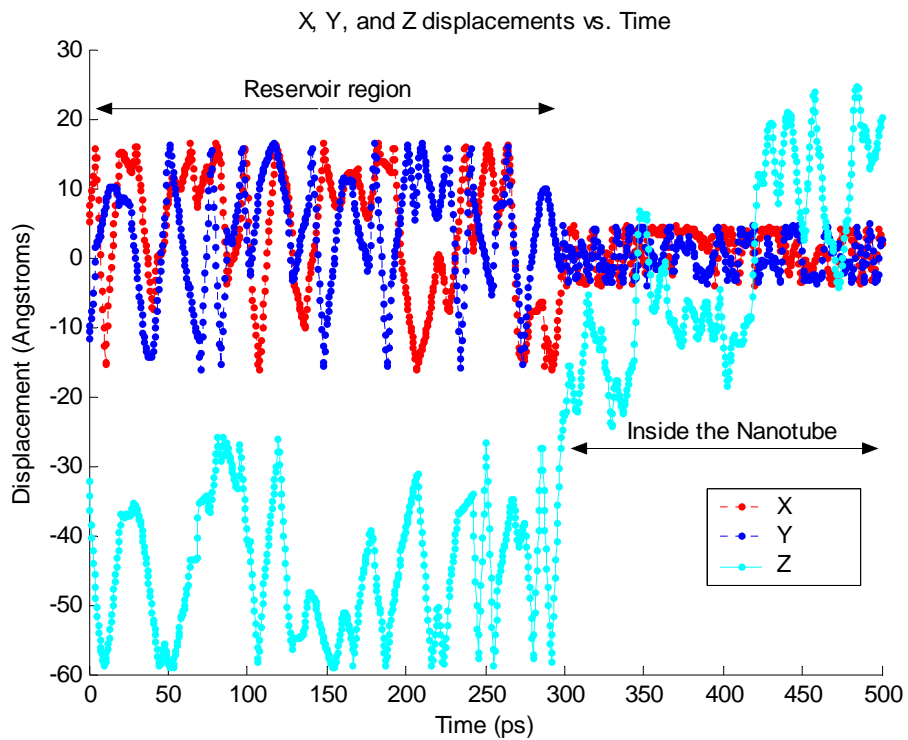

**Fig. 5.18** Displacement vs. Time plot of a random argon atom

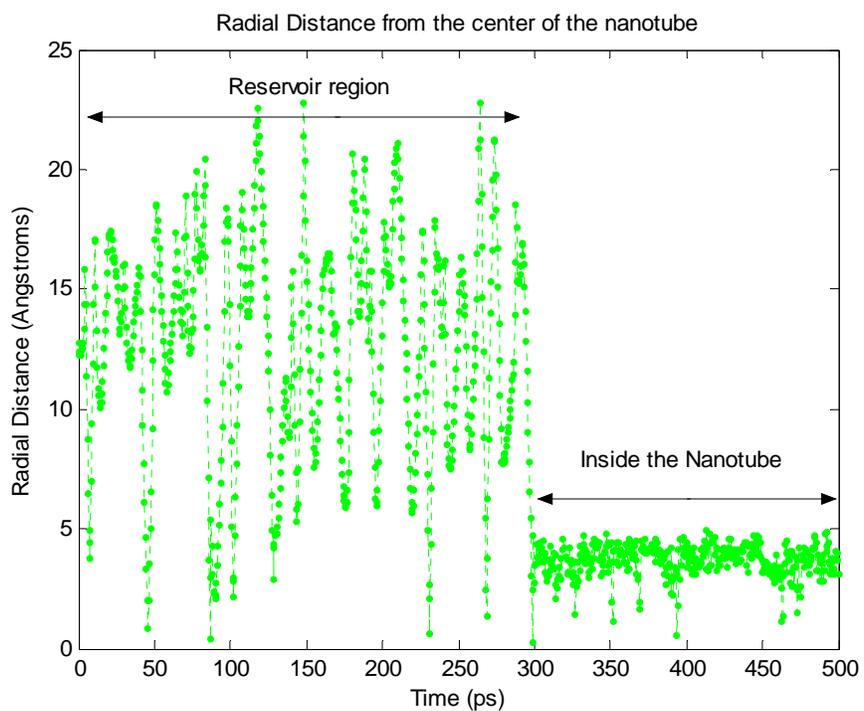

**Fig. 5.19** Radial Displacement vs. Time plot of a random argon atom

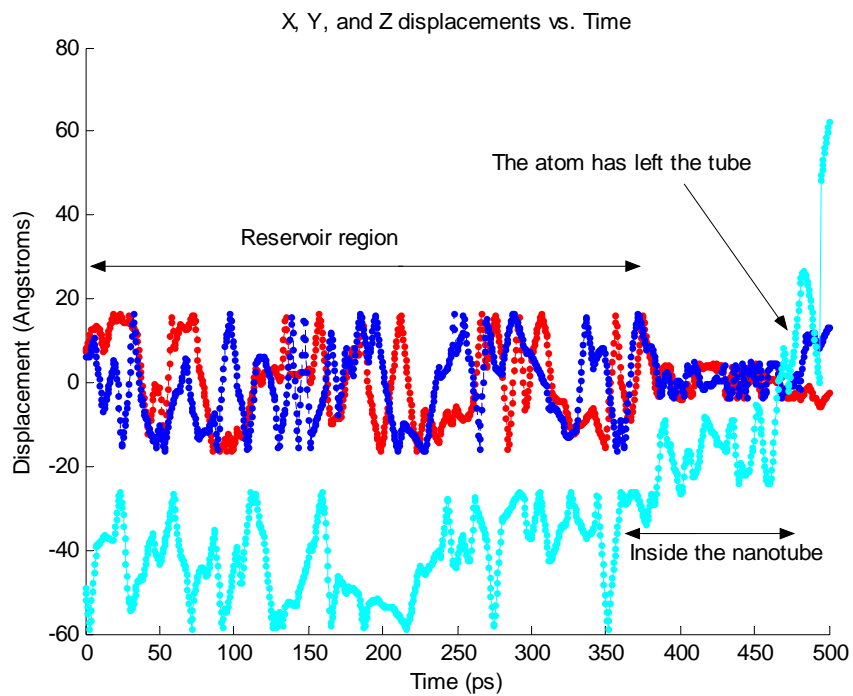

**Fig. 5.20** Displacement vs. Time plot of a second randomly selected argon atom

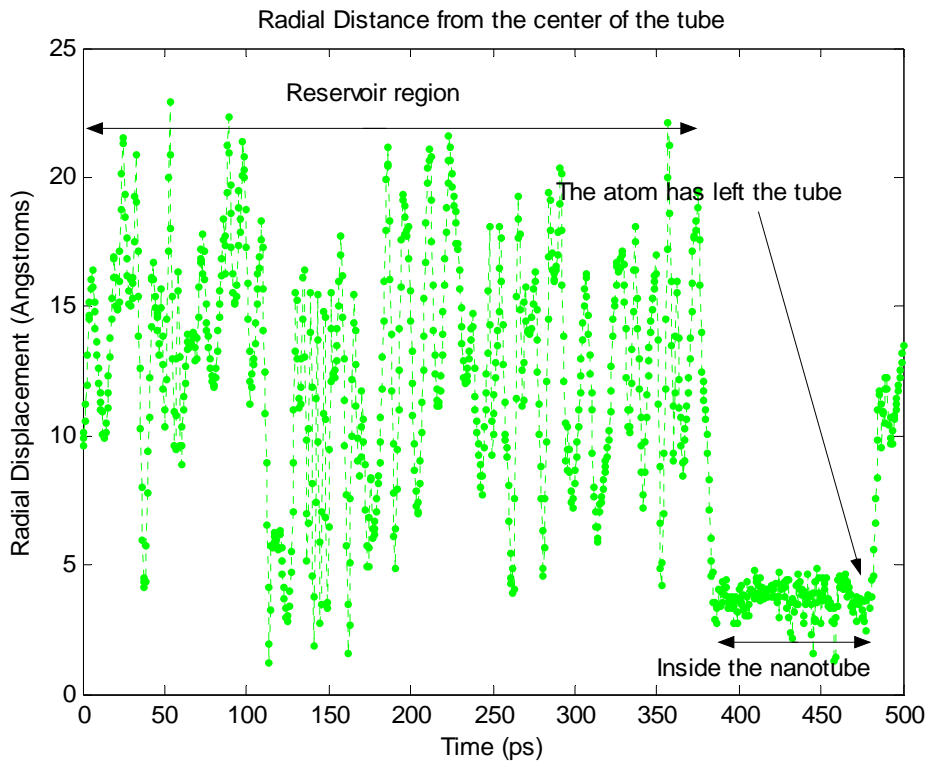

**Fig. 5.21** Radial Displacement vs. Time plot of a second randomly selected argon atom

The slope of the graph between average square displacement ($<r^2>$) and time gives the diffusion coefficient. The first and second stages are characterized by a non-linear relationship between $<r^2>$ and time. The data is more scattered as the atoms move back and forth inside the tube. It has also been suggested in literature that such an analysis to calculate the diffusion coefficient is erroneous as the flow is not steady and hence the plots are not presented here. As stated earlier the flow inside the nanotube could never reach a steady-state as the diffusion length is of the same order as the tube diameter. Hence a more detailed analysis will be needed to characterize the diffusion coefficient in entry problems such as this.

# Chapter 6
# CONCLUSIONS

This report focused on simulating gaseous argon flow, both inside and outside the nanotubes. It was observed that the flow of argon is erratic and highly random. The argon atom while traveling inside the tube takes a spiral path, thus following the shape of the nanotube. It moves close to the walls of the nanotube with sudden passes through the center. The movement of a large number of atoms inside the nanotube is highly coordinated with the atoms forming a 'spiral staircase-like' structure. Even outside the tube the atoms arrange themselves in their minimum energy positions. In order to transport fluid from one reservoir to another, a number of factors would come into play. Initially the interactions between the fluid atoms inside the reservoir should be such that the repulsions between them are greater than the repulsions with the nanotube tip atoms. Once inside the nanotube the motion of the atoms is chaotic with the atoms moving back and forth along the length of the nanotube. This happens if the kinetic energy of the atom is lesser than the attractive potential with the nanotube atoms. The curvature of the nanotube is also likely to play an important effect. As the diameter of the nanotube decreases, the effect of this potential would also increase. It is also difficult to characterize a diffusion coefficient for the flow conditions in this study as a steady-state regime was never attained.

The major idea behind this study was to understand molecular dynamics and apply it to a problem to analyze the processes happening at the nanoscale. A fundamental understanding can be attained by implementing the tools of molecular dynamics or Monte Carlo to the problems at this scale, but a complete analysis is only possible by incorporating quantum effects.

# Chapter 7

# RECOMMENDATIONS FOR FUTURE RESEARCH

The next step would be to try to simulate and characterize flows of different fluids and ions inside carbon nanotubes. Water is the most interesting fluid, as it is difficult to incorporate all inter and intra molecular interactions into one potential. Also water flow in nanotubes has a lot of applications in the fields of Biotechnology and drug delivery. The application of electric fields and its effect on the flow of polarizable and charged fluids inside nanocomponents would be investigated in future studies using molecular dynamics simulations. There is a need to understand fundamentally the development of the Electric double layer in components, which is of great significance at the nanoscale.

Anything at the nanoscale is inevitably controlled by quantum effects. In order to understand the flow of ions or other polar fluids inside nanomachines, it is important to incorporate quantum effects in the modeling. However the major disadvantage is that only a limited number of molecules can be simulated at a given time and the problem is computationally intensive. Hence future work would be based on combining these approaches by which, reasonably accurate models can be made to characterize fluid flow at the nanoscale.